\documentclass[10pt]{article}

\begin{document}

\title{Embeddings for  $4D$ Einstein equations with a cosmological constant}
\author{J. Ponce de Leon\thanks{E-Mail:
jpdel@ltp.upr.clu.edu, jpdel1@hotmail.com}  \\Laboratory of Theoretical Physics, 
Department of Physics\\ 
University of Puerto Rico, P.O. Box 23343,  
San Juan,\\ PR 00931, USA}
\date{December, 2007}

\maketitle
\begin{abstract}
There are many ways of embedding a $4D$ spacetime in a given higher-dimensional manifold while, satisfying the field equations. In this work we extend and generalize a recent paper by Mashhoon and Wesson ({\it Gen. Rel. Gravit.} {\bf 39}, 1403(2007)) by showing different ways of embedding  a solution of the $4D$ Einstein equations, in vacuum with a cosmological constant, in a Ricci-flat, as well as in an anti-de Sitter, $5D$ manifold.  These embeddings lead to  different physics in $4D$.  In particular, to non-equivalent cosmological terms as functions of the extra coordinate. We study the motion of test particles for different embeddings and show that there is a complete equivalence between several  definitions for the  effective mass of test particles measured in $4D$, obtained from different theoretical approaches like the Hamilton-Jacobi formalism and the principle of least action. For the case under consideration, we find that the effective mass observed in $4D$ is the same regardless of whether we consider null or non-null motion in $5D$.

\end{abstract}

\medskip



\newpage
\section{Introduction}

It is well known that any solution of the $4D$ Einstein equations of general relativity may be embedded in a solution of the $5D$ vacuum Einstein equations. The higher dimensional manifold is generally assumed to be either Ricci-flat   or anti-de Sitter, depending on the theory under consideration. However, in all cases finding an exact solution in $5D$ that embeds a particular metric in $4D$ is not an easy task. In spite of this, the existence of such a solution is guaranteed by Campbell-Magaard embedding theorems \cite{Rippl}-\cite{Wesson1}.

Still and all, these theorems do not show how to produce the actual embedding. In fact, there are many ways of embedding a $4D$ spacetime in a given higher-dimensional manifold while, satisfying the field equations \cite{arXiv:gr-qc/0512067}. For example, a single solution in $5D$ may generate very different scenarios in $4D$, ranging from static configurations to cosmological situations. 
Another example is provided by the multiplicity of possible non-Schwarzschild scenarios for the  effective spacetime outside of a static spherical star, admitted  in Kaluza-Klein gravity and braneworld models \cite{Cristiano}-\cite{arXiv:0711.0998}.

The crucial point is that the $(4 + 1)$ dimensional reduction of a $5D$ metric leaves  various alternatives for the construction of the metric of the physical spacetime, from the metric induced in $4D$. Since the calculation of physical effects does require the specification of a $4D$ metric, it follows that different alternatives  generally produce different physics in $4D$. Therefore, in order to be able to test the theory with observations and experiments one has to examine and work out  all the theoretical possibilities.

In a recent paper Mashhoon and Wesson \cite{Mashhoon} present an embedding for vacuum solutions of $4D$ general relativity, with cosmological constant. The embedding is a solution of the $5D$ Ricci-flat equations obtained under the following assumptions: (i) the $5D$ line element can be written in the  ``canonical" form; (ii) the  functional dependence of the metric on the extra coordinate is separable (see bellow (\ref{the metric 2})); (iii) the extra coordinate is spacelike. For this $5D$ solution, they show that the corresponding induced matter in $4D$ can be interpreted as  an effective cosmological ``constant" that is a {\it specific} function of the extra coordinate. They also discuss  various possible physical effects associated with  this interpretation.

In this work, we extend and generalize Mashhoon-Wesson's results.  Specifically, in addition to five-dimensional  Ricci-flat manifolds, we also consider anti-de Sitter ones. For these manifolds, we show that the field equations in $5D$ can be fully integrated {\it without} using the first and third assumptions mentioned above; only the second one is needed for this purpose. 

For Ricci-flat manifolds, which are relevant to theories  of Kaluza-Klein gravity \cite{Overduin}, we demonstrate that the solution of the field  equations generates three embeddings for  vacuum solutions in $4D$ general relativity, with cosmological constant. One of them is the ``canonical" embedding discussed by Mashhoon and Wesson. These embeddings lead to  different physics in $4D$.  In particular, to non-equivalent cosmological terms as functions of the extra coordinate. 

In an anti-de Sitter bulk, which is pertinent to the Randall and Sundrum braneworld scenario \cite{RS}, the solution of the field equations generates an effective cosmological term in $4D$ that, unlike the ones in Ricci-flat embeddings, can always be set equal to zero. Our analysis corroborates  that distinct scenarios in $4D$ might correspond to the same physics in $5D$.

The simplicity of the embeddings considered here provides an appropriate arena to examine various aspects of the theory. In this work we are particularly interested in testing several aspects related to  the effective mass of test particles as measured by an observer in $4D$. 
With this aim we study the motion of test particles for different embeddings  along timelike paths in $4D$, which can be massless, as well as massive  in $5D$, i.e., moving along null or non-null $5D$ geodesics. 

Firstly, our calculations show a complete equivalence between several  definitions for the  effective mass of test particles measured in $4D$, obtained from different theoretical approaches like the Hamilton-Jacobi formalism \cite {arXiv:gr-qc/0209013}, \cite{arXiv:gr-qc/0310078}  and the principle of least action \cite{arXiv:0711.1004}. For the case under consideration, we find that the effective mass observed in $4D$ is the same regardless of whether we consider null or non-null motion in $5D$.
Secondly, our results  confirm the interpretation that the anomalous acceleration, or  extra ``fifth" force, predicted by theories allowing explicit dependence of the extra dimension,   can be viewed as a $4D$ consequence of the variation of rest mass  due to their momentum along the extra dimension.  

This paper is organized as follows. In section $2$ we give a brief summary of the field equations. In sections $3$ and $4$ we integrate the equations for $5D$ Ricci-flat and anti-de Sitter manifolds, respectively,  and obtain the effective cosmological terms  for  various versions of Kaluza-Klein gravity and braneworld theory. In section $5$ we   
study geodesic motion in $5D$. For the Ricci-flat embeddings, we consider in detail null and non-null geodesic motion in $5D$. Section $6$ is a summary and discussion. In particular, following a recent study by Overduin, Wesson and Mashhoon \cite{OWM}, we compare the age of the universe for different Ricci-flat embeddings. However, a more detailed analysis is needed. This is beyond the scope of the present paper.

\section{Field equations}

In this section we give a sketch of the theoretical background needed for our discussion\footnote{In what follows $c = G = 1$; lowercase Greek letters go from $0$ to $3$; $x^{0}$ is
 timelike, $x^{1}$, $x^{2}$, $x^{3}$ are spacelike; $y$ represents the ``extra" coordinate; capital Latin letters $A, B$ denote indexes in $5D$}.  
Let us consider the $5D$ metric,
\begin{equation}
\label{the metric 1}
dS^2 = \gamma_{AB}dx^{A}d x^{B} = \gamma_{\mu \nu}(x^{\rho}, y)dx^{\mu}dx^{\nu} + \epsilon dy^2,
\end{equation}
where $y$ represents the ``extra" coordinate. The spacetime is usually assumed to be some hypersurface orthogonal to the vector field 
\begin{equation}
n_{A} = \delta_{A}^{4},
\end{equation}
along the extra dimension. Besides, $n_{A}n^{A} = \epsilon$ with $\epsilon = - 1$ or $\epsilon = 1$, depending on whether the extra dimension is spacelike or timelike, respectively. For this choice,
$\gamma_{\mu\nu}$ represents the metric {\it induced } in $4D$. 

The $(4 + 1)$ dimensional reduction of the five-dimensional Ricci tensor, which we denote as ${\bar{R}}_{AB}$, yields 
\begin{eqnarray}
\label{general expressions for bar{R}}
{\bar{R}}_{\mu \nu} &=& R_{\mu\nu} - \epsilon \left( K_{\mu\nu}' - 2K_{\mu\rho}K^{\rho}\;_{\nu} + K K_{\mu\nu}\right),\nonumber \\ 
{\bar{R}}_{44} &=& - K' - K_{\alpha \beta}K^{\alpha\beta}, \nonumber \\
{\bar{R}}_{4\mu} &=& K^{\nu}_{\mu\;;\nu} - \frac{\partial K}{\partial x^{\mu}}. 
\end{eqnarray}
Here primes denote derivative with respect to $y$; $R_{\mu\nu}$ is the Ricci tensor defined in $4D$, and $K_{\mu\nu} = g_{\mu \nu}'/2$ represents the extrinsic curvature of hypersurfaces $\Sigma: y = y_{0}$, which are orthogonal $n^A $.
Moreover  
\begin{equation}
\label{useful relations for K}
K^{\mu\nu} = - \frac{1}{2}\left(g^{\mu\nu}\right)',\;\;\;K = g_{\mu\nu}K^{\mu\nu} = g^{\lambda \rho}K_{\lambda \rho}, \;\;\;K' = g^{\mu\nu} K_{\mu\nu}' - 2K_{\mu\nu}K^{\mu\nu}.
\end{equation}
Now we consider, Mashhoon-Wessson ansatz  \cite{Mashhoon} where the induced metric is separated as
\begin{equation}
\label{the metric 2}
\gamma_{\mu \nu} (x^{\rho}, y) = \Omega(x^{\rho}, y)\sigma_{\mu\nu}(x^{\lambda}).
\end{equation}
In this case the components of the $5D$ Ricci tensor (\ref{general expressions for bar{R}}) become
\begin{eqnarray}
\label{RAB for the separated metric}
{\bar{R}}_{\mu\nu} &=& R_{\mu\nu} - \frac{\epsilon}{2}\gamma_{\mu\nu}\left(\frac{\Omega''}{\Omega} + \frac{\Omega'^{2}}{\Omega^2}\right), \nonumber \\
{\bar{R}}_{44} &=& \frac{\Omega'^2}{\Omega^2} - 2\frac{\Omega''}{\Omega}, \nonumber \\
{\bar{R}}_{\mu 4} &=& - \frac{3}{2}\frac{\partial}{\partial x^{\mu}}\left(\frac{\Omega'}{\Omega}\right).
\end{eqnarray}
We now proceed to integrate these equations for different $5D$ manifolds. 

\section{Ricci-flat manifolds: Kaluza-Klein gravity}
Many  Kaluza-Klein theories of gravity assume that the $5D$ manifold is Ricci-flat \cite{Overduin}. Therefore, we should  solve the equations
\begin{equation}
{\bar{R}}_{AB} = 0.
\end{equation}
From ${\bar{R}}_{44} = 0$ we obtain the ``trivial" solution $\Omega =$ constant and
\begin{equation}
\Omega = \left[f(x) y + h(x)\right]^2,
\end{equation}
where $f(x)$ and $h(x)$ are arbitrary functions of spacetime coordinates. One of them is fixed by ${\bar{R}}_{4\mu} = 0$. Namely, we find
\begin{equation}
\label{h in terms of f}
h(x) = C f(x),
\end{equation}
where $C$ is a constant. Thus,  
\begin{equation}
\Omega = \left[y + C\right]^2 f^{2}(x).
\end{equation} 
Now, the Ricci tensor in $4D$ becomes
\begin{equation}
\label{Ricci tensor in my calculation}
R_{\mu\nu} = 3\epsilon \sigma_{\mu\nu}f^{2}(x).
\end{equation}
Following Mashhoon and Wesson \cite {Mashhoon} we set
\begin{equation}
\label{Wesson's settings}
C = - y_{0}, \;\;\;f(x) = \frac{1}{L^2},
\end{equation}
and obtain
\begin{equation}
\label{Omega for Wesson's settings}
\Omega = \left(\frac{y - y_{0}}{L}\right)^2, \;\;\;R_{\mu\nu} = \frac{3 \epsilon}{(y - y_{0})^2}\gamma_{\mu\nu}, \;\;\;\mbox{and}\;\;\;\gamma_{\mu\nu} = \Omega \sigma_{\mu\nu}(x^{\rho}).
\end{equation}
These equations lead to an effective cosmological term for spacetime, whose explicit form depends on the interpretation of the metric in $4D$.

\subsection{STM theory}
In the original space-time-matter theory (STM), the spacetime metric $g_{\mu\nu}$ is identified with the induced one \cite{Wesson book}, i.e.,
\begin{equation}
\label{spacetime metric for STM}
dS^2 = ds^2 + \epsilon dy^2, \;\;\;\;g_{\mu\nu} \equiv \gamma_{\mu\nu}
\end{equation}
In this theory, ``ordinary" matter in $4D$ is a result of the explicit dependence  of the spacetime metric on the extra coordinate. Thus, $T_{\mu\nu}^{(ind)}$, the energy-momentum tensor (EMT) of the induced matter in $4D$ is prescribed by the  $(4 + 1)$ dimensional reduction of the $5D$ Einstein equations as 
  
\begin{equation}
\label{definition of induced matter}
8\pi T_{\mu\nu}^{(ind)} =  \epsilon \left[K_{\mu\nu}' + K \left(K_{\mu\nu} - \frac{K}{2}g_{\mu\nu}\right) - 2\left(K_{\mu\rho}K^{\rho}_{\nu} - \frac{1}{4}g_{\mu\nu} K_{\alpha \beta}K^{\alpha \beta}\right)\right].
\end{equation}
In the case under consideration
\begin{equation}
\label{case under consideration: metrica in STM}
g_{\mu\nu} = \frac{(y - y_{0})^2}{L^2}\sigma_{\mu\nu}(x).
\end{equation}
Substituting this expression into (\ref{definition of induced matter}) we obtain
\begin{equation}
\label{definition of LambdaSTM}
8\pi T_{\mu\nu}^{(ind)} =  - \frac{3 \epsilon}{(y - y_{0})^2} g_{\mu\nu}.
\end{equation}
Thus, in STM we have
\begin{equation}
\label{Lambda STM}
\Lambda_{(STM)} = - \frac{3\epsilon}{(y - y_{0})^2}.
\end{equation}

\subsection{Canonical metric}
Without entering into details, which can be found in \cite{Mashhoon}, \cite{Mashhoon2}-\cite{Seahra2}, Mashhoon and Wesson factorize the $4D$ part of the $5D$ metric by an $y^2$ term.  Consequently, the metric $g_{\mu\nu}$ is given by
\begin{equation}
\label{canonical metric}
dS^2 = \frac{y^2}{L^2} ds^2 + \epsilon dy^2, \;\;\;\mbox{that is}\;\;\;\gamma_{\mu\nu} = \frac{y^2}{L^2}g_{\mu\nu}, 
\end{equation}
where $L$ is a constant length introduced for the consistency of physical dimensions. For this metric, which is usually called canonical, the effective matter in $4D$ 
is given by  
\begin{equation}
\label{definition of induced matter for canonical metric}
8\pi T_{\mu\nu}^{(ind)} = - \frac{3 \epsilon }{L^2}g_{\mu\nu}+ \frac{\epsilon y^2}{L^2}\left[K_{\mu\nu}' + \left(K + \frac{4}{y}\right)\left(K_{\mu\nu} - \frac{K}{2}g_{\mu\nu}\right) - 2\left(K_{\mu\rho}K^{\rho}_{\nu} - \frac{1}{4}g_{\mu\nu} K_{\alpha \beta}K^{\alpha \beta}\right)\right],
\end{equation}
with  $K_{\alpha \beta} \equiv g_{\alpha \beta}'$. In the case under consideration
\begin{equation}
\label{case under consideration: canonical metric}
g_{\mu\nu} = \left(\frac{y - y_{0}}{y}\right)^2 \sigma_{\mu\nu}(x),
\end{equation}
For this specific metric, Mashhoon and Wesson (M-W) have recently shown that 
\begin{equation}
\label{Lambda M-W}
\Lambda_{(M-W)} = - \frac{3\epsilon y^2}{L^2(y - y_{0})^2}.
\end{equation}
which is exactly what we get from (\ref{definition of induced matter for canonical metric}) for the canonical metric (\ref{case under consideration: canonical metric}).

\subsection{General $\Omega$ warp factor}

A more general factorization of the metric, usually considered in the literature, is given by 
\begin{equation}
\label{general warp}
dS^2 = \Omega(x^{\rho}, y)ds^2 + \epsilon dy^2, \;\;\;\mbox{i.e., }\;\;\;     \gamma_{\mu\nu} = \Omega(x^{\rho}, y)g_{\mu\nu}(x^{\lambda}, y).
\end{equation}
The corresponding induced EMT is

\begin{equation}
\label{definition of induced matter for general warp factor}
8\pi T_{\mu\nu}^{(ind)} = - \frac{\epsilon }{2}g_{\mu\nu}\left(\frac{2 \Omega'^2}{\Omega} -  \Omega''\right) + \epsilon \Omega\left[K_{\mu\nu}' + \left(K + \frac{2 \Omega'}{\Omega}\right)\left(K_{\mu\nu} - \frac{K}{2}g_{\mu\nu}\right) - 2\left(K_{\mu\rho}K^{\rho}_{\nu} - \frac{1}{4}g_{\mu\nu} K_{\alpha \beta}K^{\alpha \beta}\right)\right].
\end{equation}
For the case under consideration, if we take 
\begin{equation}
\label{case under consideration: general Omega}
g_{\mu\nu} = \sigma_{\mu\nu}(x),
\end{equation}
then
the induced cosmological term is {\it not} a function of the extra coordinate. Namely, for  (\ref{Omega for Wesson's settings}) we get 
\begin{equation}
\label{Lambda Omega}
\Lambda_{(\Omega)} = - \frac{3 \epsilon}{L^2}.
\end{equation}
We note that for $y_{0} = 0$, $\Lambda_{(M-W)} = \Lambda_{(\Omega)} = - 3\epsilon/L^2$. For $y \gg y_{0}$ we have $\Lambda_{(M-W)} \rightarrow -3\epsilon/L^2$ and $\Lambda_{(STM)} \rightarrow 0$. Also, $\Lambda_{(STM)}$ and $\Lambda_{(M-W)}$ diverge for $y = y_{0}$.

\section{Anti-de Sitter manifolds: Braneworld models}

In braneworld models the bulk is not empty, so the field equations in $5D$ are
\begin{equation}
\label{5D equations in the bulk}
{\bar{R}}_{AB} - \frac{1}{2}\gamma_{AB}\bar{R} = k_{(5)}^2 {\bar{T}}_{AB}.
\end{equation}
Besides, the spacetime metric is taken to be the induced one, ie., $g_{\alpha\beta} \equiv \gamma_{\alpha\beta}$. As a consequence, the effective energy momentum tensor in $4D$ can be written as \cite{arXiv:gr-qc/0111011}
\begin{equation}
8 \pi T_{\alpha\beta}^{eff} = k_{(5)}^2\left[{\bar{T}}_{\alpha\beta} + g_{\alpha \beta}\left({\bar{T}}_{4}^{4} - \frac{1}{3}\bar{T}\right)\right] + 8\pi T_{\alpha\beta}^{(ind)},
\end{equation}
where $8\pi T_{\alpha\beta}^{(ind)}$ is given by (\ref{definition of induced matter}). In the Randall-Sundrum braneworld scenario our universe is identified with a singular hypersurface (called brane) embedded in a $5$-dimensional anti-de Sitter bulk with ${\bf{Z}}_{2}$ symmetry with respect to the brane. This symmetry, together with Israel's boundary conditions, yield a definite connection between the extrinsic curvature  $K_{\mu\nu}$
and the energy momentum tensor of the matter on the brane, which leads to the specific form of the EMT in braneworld theory \cite{Shiromizu}. 

Setting ${\bar{T}}_{AB} = \Lambda_{(5)}\gamma_{AB}$, from (\ref{5D equations in the bulk}) we obtain
\begin{equation}
\label{RAB for braneworld models}
{\bar{R}}_{AB} = - \frac{2}{3}k_{(5)}^2 \Lambda_{(5)}\gamma_{AB}.
\end{equation}
Again, equations  (\ref{general expressions for bar{R}}) can be integrated for the case where the braneworld metric can be separated as in (\ref{the metric 2}). In fact, substituting (\ref{RAB for braneworld models}) into (\ref{RAB for the separated metric}) we obtain an equation for $\Omega$
\begin{equation}
\label{New equation for Omega}
2\frac{\Omega''}{\Omega} - \frac{\Omega'^2}{\Omega^2} = \frac{2\epsilon}{3}k_{(5)}^2\Lambda_{(5)},
\end{equation}
whose solution is
\begin{equation}
\Omega = \left[\tilde{f}(x)e^{\omega y/2} + \tilde{h}(x)e^{- \omega y/2}\right]^2, \;\;\;\mbox{where}\;\;\;\omega \equiv \sqrt{\frac{2\epsilon}{3}k_{(5)}^2 \Lambda_{(5)}} \neq 0,
\end{equation}
and functions $\tilde{f}(x)$, as well as $\tilde{h}(x)$, are arbitrary. In anti-de Sitter manifolds $\Lambda_{(5)} < 0$, thus $\omega$ is a real number if the extra dimension is spacelike. 

Since $\gamma _{4\mu} = 0$, it follows that ${\bar{R}}_{4\mu} = 0$. As in (\ref{h in terms of f}), this equation requires $\tilde{h}(x) = \tilde{C}\tilde{f}(x)$, where $\tilde{C}$ is a dimensionless constant of integration. Substituting this expression into (\ref{RAB for the separated metric}) and using (\ref{RAB for braneworld models}) we find

\begin{equation}
R_{\mu\nu} = - \frac{3\tilde{C}\omega^2}{\left(e^{\omega y/2} + \tilde{C}e^{- \omega y/2}\right)^2}g_{\mu\nu}.
\end{equation}
Thus, in this case the effective cosmological term is given by
\begin{equation}
\Lambda_{(brane)} = \frac{3\tilde{C}\omega^2}{\left(e^{\omega y/2} + \tilde{C}e^{- \omega y/2}\right)^2}.
\end{equation}
We note that  $\Lambda_{(brane)} = 0$ for $\tilde{C} = 0$ for {\it any} choice of $y$. This should be contrasted with 
$\Lambda_{(STM)}$ and $\Lambda_{(\Omega)}$ which cannot be set equal to zero, while $\Lambda_{(M-W)} = 0$ only for the choice $y = 0$.

\section{Geodesic motion in $5D$}
The object of this section is to evaluate the anomalous acceleration, or extra force, as well as the effective mass of test particles as measured by an observer who is bounded to our $4D$ spacetime. 

Let us, therefore, consider the geodesic equation in $5D$
\begin{equation}
\label{geodesic equation in 5D}
\frac{d^2 x^A}{d\lambda^2} + \Gamma^{A}_{BC}\frac{dx^A}{d\lambda}\frac{dx^B}{d\lambda} = 0,
\end{equation}
where $\lambda$ is some affine parameter along the geodesic. The four-velocity $u^{\mu}$ of a particle is defined as
\begin{equation}
\label{definition of 4-velocity}
u^{\mu} = \frac{dx^{\mu}}{ds}, \;\;\;ds = \sqrt{g_{\alpha\beta} dx^{\alpha}dx^{\beta}},
\end{equation}
where $g_{\alpha\beta}$ is the metric of the spacetime. For the metric (\ref{the metric 1}) we find $\Gamma^{A}_{4 4} = 0$ and $\Gamma_{4\alpha}^{\mu} = (1/2)\gamma^{\mu\rho}\gamma_{\rho\alpha}'$. Therefore, the $4D$ part of (\ref{geodesic equation in 5D}) in terms of $ds$ becomes
\begin{equation}
\label{4D part of the geod. equation in terms of ds}
\frac{du^{\mu}}{ds} + \Gamma^{\mu}_{\alpha \beta}u^{\alpha}u^{\beta} = \frac{u^{\mu}}{f}\left(\frac{df}{ds}\right) - \gamma^{\mu\rho}\gamma_{\rho\lambda}' u^{\lambda}\left(\frac{dy}{ds}\right),
\end{equation}
where $f$ is  a {\it dimensionless} function defined by  
\begin{equation}
\label{definition of lambda}
d\lambda = f ds.
\end{equation}
This function is related to the extra coordinate $y$ through the fourth component of the geodesic equation (\ref{geodesic equation in 5D}). Indeed, for metric (\ref{the metric 1}) we find $\Gamma_{44}^4 = \Gamma_{\mu 4}^{4} = 0$ and $\Gamma_{\mu\nu}^4 = - (\epsilon/2)\gamma_{\mu\nu}'$. Therefore, setting $A = 4$ in (\ref{geodesic equation in 5D}) we obtain 
\begin{equation}
\label{equation for f}
\frac{d^2 y}{ds^2} - \frac{1}{f}\left(\frac{df}{ds}\right)\left(\frac{dy}{ds}\right) - \frac{\epsilon}{2}\gamma_{\mu\nu}'u^{\mu}u^{\nu} = 0.
\end{equation}

\subsection{Motion along null geodesics in $5D$}

Let us first consider that particles move along null geodesics in $5D$, i.e., $dS = 0$ along the motion. This is possible only if $\epsilon = - 1$, i.e., the extra coordinate is spacelike. 
Thus, from (\ref{spacetime metric for STM}), (\ref{canonical metric}) and (\ref{general warp}) it follows that along such geodesics $(dy/ds) = 1$;  $(dy/ds) = y/L$, and $(dy/ds)= \sqrt{\Omega}$ for the STM, canonical and $\Omega$-factor metrics, respectively. Thus, substituting into (\ref{equation for f}) and integrating we find 
\begin{equation}
\label{f for null geodesics}
f(y) = \left\{\begin{array}{cc}
            \mbox{constant} \times \frac{(y - y_{0})}{L},     & \mbox{for STM}, \\
\\
 \mbox{constant} \times y(y - y_{0}),  & \mbox{for canonical metric}, \\
\\
               \mbox{constant} \times \frac{(y - y_{0})^2}{L^2},   & \mbox{for $\Omega$ warp factor}.
               \end{array}
      \right.
\end{equation}
Substituting these expressions into (\ref{4D part of the geod. equation in terms of ds}) we obtain

\begin{equation}
\label{extra acceleration for null motion in 5D}
\frac{du^{\mu}}{ds} + \Gamma^{\mu}_{\alpha \beta}u^{\alpha}u^{\beta} = \left\{\begin{array}{cc}
                - \frac{u^{\mu}}{(y - y_{0})} & \mbox{for STM}, \\
\\
 - \frac{u^{\mu}y_{0}}{L(y - y_{0})} & \mbox{for canonical metric}, \\
\\
               0   & \mbox{for $\Omega$ warp factor}.
               \end{array}
      \right.
\end{equation}
Note that when $y_{0} = 0$, the $4D$ motion is geodesic for the canonical metric, but  not for the STM

\subsubsection{Effective mass of test particles observed in $4D$}

It should be noted that the principle of least action provides an equation for the effective mass measured by an observer in $4D$, which we denote as $m$. Indeed, equation (14) in reference \cite{arXiv:0711.1004} reads
\begin{equation}
\label{equation for m}
\frac{1}{m} \frac{\partial m}{\partial y} + \frac{1}{2}u^{\alpha}u^{\beta}\frac{\partial g_{\alpha \beta}}{\partial y} = 0,
\end{equation}
where $g_{\alpha\beta}$ is the metric of the spacetime. We also have
\begin{equation}
\label{m from the geodesic}
\frac{du_{\rho}}{ds} - \Gamma_{\rho \alpha}^{\beta} u^{\alpha}u_{\beta} = - \frac{u_{\rho}}{m}\frac{\partial m}{\partial y}\frac{dy}{ds}.
\end{equation}
Substituting here (\ref{case under consideration: metrica in STM}), (\ref{case under consideration: canonical metric}), (\ref{case under consideration: general Omega}), and integrating we obtain
\begin{equation}
\label{mass for null geodesics}
m(y) = \left\{\begin{array}{cc}
              \frac{m_{0}L}{|y - y_{0}|},     & \mbox{for STM}, \\
\\
  \frac{m_{0}y}{|y - y_{0}|},  & \mbox{for canonical metric}, \\
\\
               m_{0},   & \mbox{for $\Omega$ warp factor},
               \end{array}
      \right.
\end{equation}
where $m_{0}$ are constants of integration with the appropriate dimensions. It is important to mention that the above results can also be obtained from our previous definition of effective mass from the Hamilton-Jacobi formalism \cite{arXiv:gr-qc/0209013},  \cite{arXiv:gr-qc/0310078}. In fact, they are equivalent as shown in the conclusion section of \cite{arXiv:0711.1004}. They are also consistent with the definition of mass given by equation (44) in \cite{arXiv:0711.1004}. In order to avoid misunderstanding, let us notice that $g_{\alpha\beta} \rightarrow F g_{\alpha\beta}$ implies $ds \rightarrow \sqrt{F}ds$ and consequently $m \rightarrow \sqrt{F}/\bar{f}$ and $\bar{f} = f/\sqrt{F}$. Therefore, $m = \bar{M}/ \bar{f}$ in that paper now becomes 
\begin{equation}
\label{alternative definition of  m}
m = \frac{\bar{M}F}{f},
\end{equation} 
with $F = 1$, $F = y^2/L^2$ and $F = (y - y_{0})^2/L^2$ for  the STM, canonical and general warp factor metrics, respectively. Thus  (\ref{alternative definition of  m}) reproduces the results showed in (\ref{mass for null geodesics}).

\subsection{Non-null geodesics in $5D$}

For non-null geodesics, without loss of generality  we can take $d\lambda = dS$. Thus, 
\begin{equation}
f = \left\{\begin{array}{cc}
            \sqrt{1 + \epsilon (dy/ds)^2}    & \mbox{for STM}, \\
\\
 \sqrt{y^2/L^2 + \epsilon(dy/ds)^2}& \mbox{for canonical metric}, \\
\\
               \sqrt{(y - y_{0})^2/L^2 + \epsilon (dy/ds)^2}   & \mbox{for $\Omega$ warp factor}.
               \end{array}
      \right.
\end{equation}
Substituting into (\ref{equation for f}) and integrating we obtain
\begin{equation}
\left(\frac{dy}{ds}\right)^2 = \left\{\begin{array}{cc}
            \epsilon \left[ - 1 + C_{1}\left(y - y_{0}\right)^2\right]    & \mbox{for STM}, \\
\\
 \epsilon (y^2/L^2)\left[ - 1 + C_{2}\left(y - y_{0}\right)^2\right]& \mbox{for canonical metric}, \\
\\
   \epsilon [(y-y_{0})/L]^2\left[ - 1 + C_{3}\left(y - y_{0}\right)^2\right]               & \mbox{for $\Omega$ warp factor},
               \end{array}
      \right.
\end{equation}
where $C_{1}$, $C_{2}$ and $C_{3}$ are constants of integration. Thus, for non-null geodesics instead of (\ref{extra acceleration for null motion in 5D}) we have
\begin{equation}
\label{extra acceleration for non-null motion in 5D}
\frac{du^{\mu}}{ds} + \Gamma^{\mu}_{\alpha \beta}u^{\alpha}u^{\beta} = \left\{\begin{array}{cc}
            - \frac{u^{\mu}}{y - y_{0}}\left(\frac{dy}{ds}\right)    & \mbox{for STM}, \\
\\
 - \frac{y_{0}}{y(y - y_{0})}\left(\frac{dy}{ds}\right)  &\mbox{for canonical metric}, \\
\\
               0  & \mbox{for $\Omega$ warp factor}.
               \end{array}
      \right.
\end{equation}
We note that setting $C_{1} = C_{2} = 0$, and $\epsilon = - 1$ we recover the results for null geodesics (\ref{extra acceleration for null motion in 5D}). In addition, we find total consistency between various definitions  for the  effective mass of test particles observed in $4D$. Namely, $m$ calculated from (\ref{equation for m}), (\ref{m from the geodesic}),  the Hamilton-Jakobi formalism \cite{arXiv:gr-qc/0209013}, \cite{arXiv:gr-qc/0310078},  and other equations provided in \cite{arXiv:0711.1004} lead to the same result, which is identical to the one calculated for null geodesics (\ref{mass for null geodesics}).

Thus, for the case under consideration  the effective mass $m$ measured in $4D$ does {\it not} depend on whether we are assuming null or non-null geodesics motion in $5D$. 

\section{Summary and conclusions}
We have discussed various embeddings of solutions of the $4D$ Einstein equations, in vacuum with a cosmological constant, in Ricci-flat and anti-de Sitter manifolds. We have seen  that the effective cosmological term and the effective mass of test particles generally depend on the extra  coordinate.

Therefore, in order to have explicit expressions for these quantities one has to solve the geodesic equation to obtain  
\begin{equation}
y = y(s), \;\;\;\mbox{and}\;\;\; x^{\mu} = x^{\mu}(s).
\end{equation}
In practice this means they vary with time. Indeed, inverting the relation $t = t(s)$ to $s = s(t)$ we have $y = y(s) = y(s(t)) = y(t)$.

As an illustration, let us consider the Ricci-flat $5D$ metric

\begin{equation}
\label{metric for the example}
dS^2 = \frac{(y - y_{0})^2}{L^2}\left\{dt^2 - e^{2t/L}\left[dr^2 + r^2 (d\theta^2 + \sin^2 \theta d\phi^2)\right]\right \} - dy^2.
\end{equation}
The line element inside the curly brackets is the usual de Sitter solution of $4D$ general relativity with cosmological constant $3/L^2$. For the sake of argument, let us assume that the observer who is measuring the cosmological term is at rest in space and moves along a null geodesic in $5D$, i.e.,
 
\begin{equation}
dx^{i} = 0, \;\;\;\mbox{and}\;\;\;dS = 0. 
\end{equation}
Consequently, along its motion
\begin{equation}
\label{y as a function of t}
y = y_{0} + A e^{\pm (t/L)},
\end{equation}
 where $A$ is a constant of integration with the appropriate units. If we choose the positive sign, then $\Lambda_{(STM)}$ and $\Lambda_{(M-W)}$ are unbounded at the big bang, which in our toy model (\ref{metric for the example}) is  $t \rightarrow - \infty$. Then, they exponentially decay to 
$\Lambda_{(STM)} \rightarrow 0$ and $\Lambda_{(M-W)} \rightarrow 3/L^2$ in cosmological time.

In order to make contact with a recent study by Overduin, Wesson and Mashhoon \cite{OWM}, let us consider the expressions for $\Lambda$ in more detail. Substituting (\ref{y as a function of t}) into (\ref{Lambda STM}) and (\ref{Lambda M-W}) we obtain
\begin{equation}
\label{Lambda as a function of time}
\Lambda_{(STM)} = \frac{3\alpha^2}{y_{0}^2}e^{- 2t/L}, \;\;\;
\Lambda_{(M-W)} = \frac{3}{L^2}\left[1 \pm \alpha e^{- (t/L)}\right]^2, \;\;\;\mbox{where}\;\;\;\pm \alpha \equiv \frac{y_{0}}{A}.
\end{equation}
In this notation, the expression for $\Lambda_{(M-W)}$ is identical to the one used by these authors in \cite{OWM}, namely equation $(6)$ in that paper, which is the starting-point of their investigation. Evaluating this expression at the present time $t_{0}$; restoring physical units; using the critical density $\Omega_{\Lambda,0}$ and the dimensionless quantities
\begin{equation}
\label{dimensionless quantities}
\tau \equiv H_{0}t, \;\;\;{\cal{L}} \equiv H_{0}L/c,
\end{equation}
where $H_{0}$ is the present value of Hubble's parameter, they obtain the age of the universe $\tau_{0} \equiv H_{0}t_{0}$ as 
\begin{equation}
\label{age of the universe for Lambda M-W}
{\tau_{0}}_{(M-W)} = {\cal{L}}\ln{\left(\frac{\pm \alpha}{{\cal{L}}\sqrt{\Omega_{\Lambda,0}}- 1}\right)}.
\end{equation} 
Now, applying the same procedure for $\Lambda_{(STM)}$, we obtain\footnote{Since, according to (\ref{Wesson's settings}),  $y_{0}$ is a constant in what follows we set $y_{0} = L$.} 
\begin{equation}
\label{age of the universe for STM}
{\tau_{0}}_{(STM)} = {\cal{L}}\ln{\left(\frac{\pm \alpha}{{\cal{L}}\sqrt{\Omega_{\Lambda, 0}}}\right)}.
\end{equation}
Thus,
\begin{equation}
\label{difference of ages}
{\tau_{0}}_{(M-W)} - {\tau_{0}}_{(STM)} = {\cal{L}}\ln{\left(\frac{{\cal{L}}\sqrt{\Omega_{\Lambda,0}}}{|{\cal{L}}\sqrt{\Omega_{\Lambda,0}} - 1|}\right)}.
\end{equation}
Consequently, a universe with a cosmological term $\Lambda_{(M-W)}$ is {\it older} than the one with $\Lambda_{(STM)}$, i.e.,  
\begin{equation}
{\tau_{0}}_{(M-W)} > {\tau_{0}}_{(STM)}, 
\end{equation}
for (i) $\alpha > 0$; ${\cal{L}}\sqrt{\Omega_{\Lambda,0}} > 1$, i.e.,  $L > L_{crit}$,
and (ii) $\alpha < 0$; $1/2 <{\cal{L}}\sqrt{\Omega_{\Lambda,0}} < 1$, i.e., $L_{crit}/2 < L < L_{crit}$, where $L_{crit}$ is just the de Sitter radius of standard cosmology, $L_{crit} = {c}/({ H_{0}\sqrt{\Omega_{\Lambda, 0}}})$, which takes the value of $L_{crit} = 4.9$ Gpc \cite {OWM} for  WMAP values of $H_{0}$ and $\Omega_{\Lambda,0}$ \cite{Spergel}. 

Clearly, ${\tau_{0}}_{(M-W)} \leq {\tau_{0}}_{(STM)}$, only for $\alpha < 0$ and $L \leq L_{crit}/2$.
We note that $L = L_{crit}$ corresponds to the case of constant $\Lambda$ given by (\ref{Lambda Omega}). Therefore, the denominators in (\ref{age of the universe for Lambda M-W}) and (\ref{difference of ages}) never become zero. In other words $L \rightarrow L_{crit}$ is a limiting case \cite{OWM} giving back the embedding with constant cosmological term (\ref{Lambda Omega}). A more detailed investigation of the physical consequences of the embeddings under consideration would take us  far beyond the scope of this work.  

The results of this work evidence that even in the simplest case, where the conformal factor $\Omega$ in
\begin{equation}
\gamma_{\mu\nu} = \Omega(x^{\lambda}, y)g_{\mu\nu}(x^{\rho}, y)
\end{equation}
is a function  only of $y$,  the physics in $4D$, calculated with the spacetime metric $g_{\mu\nu}$, may crucially depend on this factor.

\paragraph{Acknowledgments:}I would like to thank Bahram Mashhoon and Paul S. Wesson  for their comments on the first version of this paper.


\begin{thebibliography}{99}
\bibitem{Rippl}{S. Rippl, C. Romero and R. Tavakol, {\em Class.Quant.Grav.} {\bf 12},  2411(1995); arXiv:gr-qc/9511016.}

\bibitem{Lidsey}{J. E. Lidsey, C. Romero, R. Tavakol and S. Rippl, {\em Class.Quant.Grav.} {\bf 14}, 865(1997); arXiv:gr-qc/9907040.}


\bibitem{Dahia}{F. Dahia and  C. Romero, {\em J.Math.Phys.} {\bf 43}, 5804(2002); arXiv:gr-qc/0109076.}


\bibitem{Seahra}{S.S. Seahra and P.S. Wesson, {\em Class.Quant.Grav.} {\bf 20}, 1321(2003); arXiv:gr-qc/0302015.}

\bibitem{Wesson1}{P. S. Wesson, {\em In Defense of Campbell's Theorum as a Frame for New Physics}, arXiv:gr-qc/0507107. }

\bibitem{arXiv:gr-qc/0512067}{ J. Ponce de Leon, {\em Class.Quant.Grav.} {\bf 23}, 3043(2006); arXiv:gr-qc/0512067.}

\bibitem{Cristiano}{C. Germani and  R. Maartens, {\em Phys. Rev.} {\bf D64}, 124010(2001); hep-th/0107011.}
\bibitem{Bruni}{M. Bruni, C. Germani and R. Maartens, {\em Phys. Rev. Lett.}
{\bf 87}, 231302(2001);   gr-qc/0108013.}
\bibitem{Kofinas}{G. Kofinas and E. Papantonopoulos, {\em J. Cosmol. Astropart. Phys.} {\bf 12},  11(2004); gr-qc/0401047.}
\bibitem{Dadhich}{N. Dadhich, R. Maartens, P. Papadopoulos and V. Rezania, {\em Phys.Lett.} {\bf B487},  1(2000); 
hep-th/0003061v3.}
\bibitem{Casadio}{R. Casadio, A. Fabbri and L. Mazzacurati, {\em Phys.Rev.} {\bf D65}, 084040(2002);  
gr-qc/0111072v2.}
\bibitem{Viser}{M. Visser and D. L. Wiltshire, {\em Phys.Rev.} {\bf D67}, 104004(2003); hep-th/0212333v2.}
\bibitem{Bronnikov}{K.A. Bronnikov, H. Dehnen and V.N. Melnikov, {\em Phys.Rev.} {\bf D68},   024025(2003); gr-qc/0304068v1.}
\bibitem{arXiv:0711.0998}{J. Ponce de Leon, {\em Stellar models with Schwarzschild and non-Schwarzschild vacuum exteriors
} To appear in Gravitation and Cosmology (2008); arXiv:0711.0998. }

\bibitem{Mashhoon}{B. Mashhoon and P.Wesson, {\em General Relativity and Gravitation} {\bf 39}, 1403(2007); arXiv:0705.0067. }

\bibitem{Overduin}{J. M. Overduin and P. S. Wesson, {\em Phys.Rept.} {\bf 283}, 303(1997); 
 arXiv:gr-qc/9805018.}

\bibitem{RS}{L. Randall and S. Sundrum, {\em Phys. Rev. Lett.} {\bf{83}}, 4690(1999); arXiv:hep-th/9906064.}

\bibitem{arXiv:gr-qc/0209013}{ J. Ponce de Leon, {\em Int.J.Mod.Phys.} {\bf D12}, 757(2003); arXiv:gr-qc/0209013.}

\bibitem{arXiv:gr-qc/0310078}{J. Ponce de Leon, {\em Gen.Rel.Grav.} {\bf 36},  1333(2004); arXiv:gr-qc/0310078.}


\bibitem{arXiv:0711.1004}{J. Ponce de Leon, {\em The principle of least action for test particles in a four-dimensional spacetime embedded in 5D}; arXiv:0711.1004.}

\bibitem{OWM}{J.M. Overduin, P.S. Wesson and B. Mashhoon, {\em Astronomy $\&$ Astrophysics} {\bf 473}, 727(2007); arXiv:0707.3148.}


\bibitem{Wesson book}{P.S. Wesson, {\em Space-Time-Matter} (World Scientific Publishing Co. Pte. Ltd. 1999).}

\bibitem{Mashhoon2}{B. Mashhoon, H. Liu and P.S. Wesson, {\em Phys. Lett.} {\bf 331}, 305(1994).}

\bibitem{Wesson3}{P.S. Wesson, {\em J.Math.Phys.} {\bf 43}, 2423(2002); arXiv:gr-qc/0105059. }

\bibitem{Wesson4}{B. Mashhoon and P. S. Wesson, {\em Class.Quant.Grav.} {\bf 21}, 3611(2004); 
 arXiv:gr-qc/0401002. }

\bibitem{Seahra2}{S.S. Seahra and P.S. Wesson, {\em Gen.Rel.Grav.} {\bf 33},  1731(2001); arXiv:gr-qc/0105041. }

\bibitem{arXiv:gr-qc/0111011}{J. Ponce de Leon, {\em Mod.Phys.Lett.} {\bf A16}, 2291(2001); 
arXiv:gr-qc/0111011.}


\bibitem{Shiromizu}{T. Shiromizu, K. Maeda and M. Sasaki, {\em Phys.Rev.} {\bf D62}, 024012(2000); arXiv:gr-qc/9910076.}

\bibitem{Spergel}{H.V.Peiris, E.Komatsu, L.Verde, D.N.Spergel, C.L.Bennett, M.Halpern, G.Hinshaw, N.Jarosik, A.Kogut, M.Limon, S.Meyer, L.Page, G.S.Tucker, E.Wollack and  E.L.Wright, {\em Astrophys.J.Suppl.} {\bf 148}, 213(2003); arXiv:astro-ph/0302225.}



\end{thebibliography}
\end{document}